\documentclass[epj]{svjour}
\usepackage{epsfig}
\def\be{\begin{equation}}
\def\ee{\end{equation}}
\def\bea{\begin{eqnarray}}
\def\eea{\end{eqnarray}}
\def\ot{(1\leftrightarrow 2)}
\def\Journal#1#2#3#4{{#1} {\bf #2}, (#4) #3 }
\def\NPA{{\em Nucl. Phys.} A}
\def\NPB{{\em Nucl. Phys.} B}
\def\PRL{\em Phys. Rev. Lett.}
\def\PRC{{\em Phys. Rev.} C}

\def\PRW{\em Phys. Rev.}

\def\PLB{{\em Phys. Lett.} B}
\def\PR{\em Phys. Rep.}
\def\IJMPA{{\em Int. J. Mod. Phys.} A}

\def\CPC{\em Comp. Phys. Comm.\,\,}
\def\ANP{\em Adv. Nucl. Phys.\,\,}
\def\JPG{{\em J. Phys.}  G}
\def\AJ{\em Astrophys. J.\,\,}

\begin{document}
\title{Interactions of the solar neutrinos with the
deuterons}
\author{B.~Mosconi\inst{1} \and P.~Ricci\inst{2}
\and \underline{E.~Truhl\'{\i}k}\inst{3}
\thanks{Presented at the International Conference 'Nuclear
Physics in Astrophysics', Debrecen, Hungary, 16th-20th May, 2005.}
}                     
\institute{Universit$\grave{a}$ di Firenze, Department of Physics,
and Istituto Nazionale di Fisica Nucleare, Sezione di Firenze,
I-50019, Sesto Fiorentino (Firenze), Italy, \email{
mosconi@fi.infn.it} \and Istituto Nazionale di Fisica Nucleare,
Sezione di Firenze, I-50019, Sesto Fiorentino (Firenze), Italy,
\email{ ricci@fi.infn.it} \and Institute of Nuclear Physics,
Academy of Sciences of the Czech Republic, CZ--250 68 \v{R}e\v{z},
Czechia, \email{ truhlik@ujf.cas.cz}}
\date{Received: date / Revised version: date}
%
\abstract{ Starting from chiral Lagrangians, possessing the
$SU(2)_L\times SU(2)_R$ local chiral symmetry,  we derive  weak
axial one--boson exchange currents in the leading order in the
$1/M$ expansion ($M$ is the nucleon mass), suitable for the nuclear
physics calculations beyond the threshold energies and with the
wave functions, obtained by solving the Schr\"odinger equation
with the one--boson exchange potentials. The constructed currents
obey the nuclear form of the partial conservation of the axial
current. We apply the space component of these currents in
calculations of the cross sections for the disintegration of
deuterons by the low energy neutrinos. The deuteron and the
$^{1}S_0$ final state nucleon--nucleon wave functions are derived
(i) from a variant of the OBEPQB potential and (ii) from the
Nijmegen 93 and Nijmegen I nucleon-nucleon interactions. The
extracted values of the constant $L_{1,\,A}$, entering the axial
exchange currents of the pionless effective field theory (EFT), are in
agreement with those predicted by the dimensional analysis.
The comparison of our cross sections with those obtained within
the pionless EFT and other potential model
calculations shows that the solar neutrino--deuteron cross
sections can be calculated within an accuracy of $\approx$ 3.3 \%.
%
\PACS{
      {11.40.Ha}{Partially conserved axial-vector
     currents}   \and
      {25.30.-c}{Lepton-induced nuclear reactions}
     } 
} 
\maketitle
\section{Introduction}\label{intro}
The semileptonic weak nuclear interaction has been studied for
half a century. The cornestones of this field of research are (i)
the chiral symmetry, (ii) the conserved vector current and (iii)
the partial conservation of the axial current (PCAC). In the
formulation \cite{SLA}, the PCAC reads
\be q_\mu\,<\Psi_f|j^a_{5\mu}(q)|\Psi_i>\,=\,if_\pi
 m^2_\pi\Delta^\pi_F(q^2)\, <\Psi_f|m^a_\pi(q)|\Psi_i>\,,
 \label{PCAC}
\ee where $j^a_{5\mu}(q)$ is the total weak axial isovector hadron
current, $m^a_\pi(q)$ is the pion source (the pion
production/absor- ption amplitude) and $|\Psi_{i,f}>$ is the wave
function describing the initial (i) or final (f) nuclear state. It
has been recognized \cite{BS} in studying the triton beta
decay
\be ^{3}H\,\rightarrow\,^{3}He\,+\,e^-\,+\,{\bar \nu}\,,
\label{TBD} \ee
and the muon capture \cite{DFM}
\be
\mu^-\,+\,^{3}He\,\rightarrow\,^{3}H\,+\,\nu_\mu\,, \label{MCHe}
\ee \be \mu^-\,+\,d\,\rightarrow\,n\,+\,n\,+\,\nu_\mu\,, \label{MCD}
\ee
that  in addition to the one--nucleon current,
the effect of the space component of weak axial exchange currents
(WAECs) enhances sensibly the Gamow--Teller matrix elements
entering the transition rates. This suggests that the current
$j^a_{5\mu}(q)$ can be understood for the system of $A$ nucleons
as  the sum of the one- and two--nucleon components,
\be
j^a_{5\mu}(q)\,=\,\sum^A_{i=1}\,j^{a}_{5\mu}(1,i,q_i)\,+\,\sum^A_{i<j}\,
j^{a}_{5\mu}(2,ij,q)\,.  \label{jtot} \ee
\par
Let us describe the nuclear system by  the Schr\"odinger equation
\be
 H|\Psi>\,=\,E|\Psi>\,,\quad  H\,=\,T\,+\,V\,, \label{NEM}
\ee
where $H$ is the nuclear Hamiltonian, $T$ is the kinetic
energy and $V$ is the nuclear potential describing the interaction
between nucleon pairs. Taking for simplicity $A=2$, we obtain from
Eq.\,(\ref{PCAC}) in the operator form and from Eqs.\,(\ref{jtot})
and (\ref{NEM}) the following set of equations for the one- and
two--nucleon components of the total axial current
\bea \vec q_i \cdot \vec j^a_{\,5}(1,\vec
q_i)\,&=&\,[\,T_i\,,\,\rho^a_{\,5} (1,\vec q_i)\,]\,+\,if_\pi
m^2_\pi \Delta^\pi_F(q^2)\nonumber \\
& & \,\times\,m^a_\pi(1,\vec q_i)
\,,\quad i=1,2\,,  \label{NCEoi}  \\
\vec q \cdot \vec j^a_{\,5}(2,\vec q)\,&=&\,[\,T_1+T_2\,,\,
\rho^a_{\,5}(2,\vec q)\,]\,+\,([\,V\,,\,\rho^a_{\,5}(1,\vec q)\,]
\nonumber \\
& & +\ot)+if_\pi m^2_\pi \Delta^\pi_F(q^2)m^a_\pi(2,\vec q).
\label{NCEt} \eea
\par
If the WAECs are constructed so that they satisfy
Eq.\,({\ref{NCEt}), then the matrix element of the total current,
sandwiched between solutions of the nuclear equation of motion
(\ref{NEM}), satisfies the PCAC (\ref{PCAC}).
\par
It is known from the dimensional analysis \cite{KDR}, that the
space component of the WAECs,  $\vec j^a_{\,5}(2,\vec q)$, is of the
order ${\cal O}(1/M^3)$. Being of a relativistic origin, it is
model dependent. This component of the WAECs was derived by
several authors in various models. In the standard nuclear physics
approach \cite{IT,ISTPR,Sci,TR,SMA}, the model systems of strongly
interacting particles contain various particles (effective degrees
of freedom), such as $N$, $\Delta(1236)$, $\pi$, $\rho$, $\omega$
and other baryons and mesons. Using these effective degrees of
freedom and chiral Lagrangians, it was possible to describe
reasonably nuclear electroweak phenomena in the whole region of
intermediate energies. In particular, the existence of  mesonic
degrees of freedom in nuclei, manifesting themselves via  meson
exchange currents, was proven to a high degree of reliability
\cite{DFM}.
\par
One of the employed Lagrangians is the one \cite{IT} containing
the heavy meson fields $\rho$ and $a_1$, taken as the Yang--Mills
gauge fields \cite{YM}. It reflects the $SU(2)_L\times SU(2)_R$
local chiral symmetry. Another used Lagrangian has been built up
\cite{STG} within the  concept of hidden local symmetries
\cite{HLSLM,BKY}. Besides possessing the chiral symmetry, our
Lagrangians are characterized by the following properties: (i)
They respect the vector dominance model, reproduce universality,
KSFRI, FSFR2, (ii) they provide correct anomalous magnetic moment
of the $a_1$ meson (iii) at the tree--level approximation, they
correctly describe elementary processes in the whole region of
intermediate energies ($E\,<\,1$ GeV) and (iv) the current algebra
prediction for the weak pion production amplitude is reproduced.
Using such an approach, the exchange currents are constructed as
follows. First, one derives the exchange amplitudes
$J^a_{5\mu}(2)$ as Feynman tree graphs. These amplitudes satisfy
the PCAC equation
\be
q_\mu\,
J^a_{5\mu}(2)\,=\,if_\pi\,m^2_\pi\,\Delta^\pi_F(q^2)\, M^a(2)\,,
\label{rPCAC}
\ee
where $M^a(2)$ are the associated pion
absorption/production amplitudes\footnote{We refer the reader for
more details to Ref.\,\cite{MRT}.}. The nuclear exchange currents
are constructed from these amplitudes in conjunction with the
equation, describing the nuclear states. Such exchange currents,
combined with the one-nucleon currents, should satisfy
Eq.(\ref{PCAC}). In the present case, we describe the nuclear
system by the Hamiltonian $H=T+V$ and the nuclear states by the
Schr\"odinger equation (\ref{NEM}). The nuclear exchange currents
are constructed within the extended S-matrix method, in analogy
with the electromagnetic meson exchange currents \cite{ATA}, as
the difference between the relativistic amplitudes $J^a_{5\mu}(2)$
and the first Born iteration of the weak axial one--nucleon
current contribution to the two--nucleon scattering amplitude,
satisfying the Lippmann--Schwinger equation. This method has
already been applied \cite{AHHST,TK1} to construct the space
component of the WAECs of the pion range.
\par
On the other hand, effective fields theories (EFTs) are being
developed since early 90's. In this approach, one starts from a
general chiral invariant Lagrangian with heavy particle degrees of
freedom integrated out and preserving $N$, $\Delta(1232)$ and
$\pi$ \cite{HHK}, or $N$ and $\pi$ \cite{PMR,PKMR}, or only
nucleons \cite{KSW,CRS}. Such EFTs rely on systematic counting rules
and on the existence of an expansion
parameter,  governing a perturbation scheme that converges
reasonably fast. The expansion parameter is given as the ratio of
the light and heavy scales.
\par
In the pionless EFT \cite{KSW,CRS}, the heavy scale $\Lambda$ is
set to the pion mass $m_\pi$. This choice restricts the
application of the scheme to the processes taking place at
threshold energies, such as the interaction of solar neutrinos
with the deuterons \cite{MB2}. In the EFT with pions, the heavy
scale is $\Lambda\,\approx\,4\pi f_\pi\,\approx\,1$ GeV,
restricting the application of the EFT to low energies.
\par
The  goal of this study is twofold: (i) The construction of the
WAECs of the heavy meson range, suitable in the standard nuclear
physics calculations beyond the long--wave limit, with the nuclear
wave functions generated from the Schr\"odinger equation using the
one--boson exchange potentials (OBEPs).  (ii)
An application of the developed formalism to the description of
the interaction of the low energy neutrinos with the deuterons,
\bea
\nu_x\,+\,d\,&\longrightarrow&\,\nu'_x\,+\,n\,+\,p\,, \label{NCN} \\
\overline{\nu}_x\,+\,d\,&\longrightarrow&\,\overline{\nu}'_x\,+\,n\,+\,p\,, \label{NCA} \\
\nu_e\,+\,d\,&\longrightarrow&\,e^-\,+\,p\,+\,p\,,  \label{CCN} \\
\overline{\nu}_e\,+\,d\,&\longrightarrow&\,e^+\,+\,n\,+\,n\,.
\label{CCA} \eea where $\nu_x$ refers to any active flavor of the
neutrino. The reactions (\ref{NCN}) and (\ref{CCN}) are important
for studying the solar neutrino oscillations, whereas the
reactions (\ref{NCA}) and (\ref{CCA}) occur in experiments with
reactor antineutrino beams. The cross sections for the  reactions
(\ref{NCN}) and (\ref{CCN}) are important for the analysis of the
results obtained in the SNO detector \cite{SNO1,SNO2,SNO4}.
The standard nuclear physics calculations \cite{NSGK,YHH}
generally differ \cite{MB2} by 5\%-10\%, which
provides a good motivation to make independent calculations
aiming to reduce this uncertainty.
\par
In Ref.\,\cite{MB2}, the effective cross sections for the
reactions (\ref{NCN})-(\ref{CCA}) are  presented in the form \be
\sigma_{EFT}(E_\nu)\,=\,a(E_\nu)\,+\,L_{1,\,A}\,b(E_\nu)\,.
\label{sigB} \ee The amplitudes $a(E_\nu)$ and $b(E_\nu)$ are
tabulated in \cite{MB2} for each of the reactions
(\ref{NCN})--(\ref{CCA}) from the lowest possible  (anti)neutrino
energy up to 20 MeV, with 1 MeV step. The constant $L_{1,\,A}$
cannot be determined from reactions between elementary particles.
Here we extract $L_{1,\,A}$ from our cross sections calculated in
the approximations of \cite{MB2}:
only the $^{1}S_0$ wave is taken into account in the nucleon--nucleon
final state and the nucleon variables are treated
non-relativistically. The knowledge of $L_{1,\,A}$ allows us to
compare our cross sections with $\sigma_{EFT}(E_\nu)$.

\section{Weak axial nuclear exchange currents}
\label{sec:1}

The starting quantities of our construction are the relativistic
Feynman amplitudes $J^a_{5\mu,B}(2)$ of the range $B$ ($B$=$\pi$,
$\rho$, $\omega$, $a_1$). These amplitudes satisfy the PCAC
constraint (\ref{rPCAC}). The WAECs $j^a_{5\mu,\,B}(2)$ of the
range $B$ are defined as \cite{MRT} \be
j^a_{5\mu,\,B}(2)\,=\,J^a_{5\mu,\,B}(2)\,-\,t^{a,\,FBI}_{5\mu,\,B}\,,
\label{JnaB} \ee where $t^{a,\,FBI}_{5\mu,\,B}$ is the first Born
iteration of the one--nucleon current contribution  to the
two--nucleon scattering amplitude, satisfying the
Lippmann--Schwinger equation \cite{ATA}.
\par
\begin{figure}
\centerline{\epsfig{file=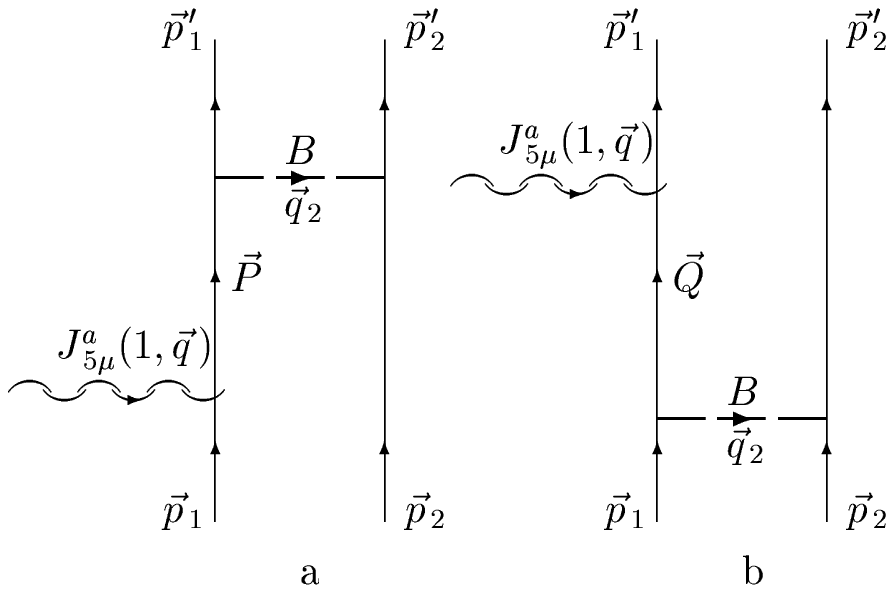} }
\vskip 0.4cm
\caption{The kinematics of the first Born iteration.
The nucleon line in the intermediate state is on--shell. }
\label{figg2}
\end{figure}
The PCAC for the WAECs, defined in Eq.\,(\ref{JnaB}), is given by
 \bea q_\mu
j^a_{5\mu,\,B}(2)\,&=&\,\left(\left[V_B,\rho^a_5(1)\right]\,
 +\,\ot \right)   \nonumber \\
& &   +\,
i f_\pi m^2_\pi \Delta^\pi_F(\vec
q^{\,2}){m}^a_B(2)\, \,, \label{dWANEC} \eea where
the nuclear pion production/absorption amplitude is given by \be
{m}^a_B(2)\,=\,M^a_B(2)\,-\,m^{a,\,FBI}_B\,,  \label{bMaB}
\end{equation}
$V_B$ is the potential of the range $B$ and $\rho^a_5(1)$ is the
one--nucleon axial charge density. We note here that the continuity equation
(\ref{dWANEC}) for our WAECs coincides with Eq.\,(\ref{NCEt}).
\par It follows from Eq.\,(\ref{dWANEC}) that in order to make
consistent calculations of the exchange current effects, one
should use OBEPs for the generation
of the nuclear wave functions and apply in the WAECs the same
couplings and strong form factors as in the potentials. In our
calculations, we employ the realistic OBE potentials OBEPQG
\cite{OPT}, Nijmegen 93 (Nijm93) and Nijmegen I (NijmI) \cite{SKTS}.
The potential
OBEPQG is the potential OBEPQB \cite{Mac}, extended by including
the $a_1$ exchange. The potential NijmI is the high quality
second generation potential with the $\chi^2$/data=1.03.
\par
In the next section, we use the WAECs, derived in the chiral
invariant models \cite{MRT,TK1,CT}, to calculate the cross
sections for the reactions (\ref{NCN})-(\ref{CCA}). By
comparing them with the EFT cross sections (\ref{sigB}),
we extract the value of the
constant $L_{1,\,A}$. We also compare our cross sections with
the cross sections of Refs.\,\cite{NSGK,YHH}. Our WAECs contain
the following components \cite{MRT}: the pair terms $\vec
j^{\,\,a}_{\,5,\,B}(pair)$ ($B=\pi$, $\rho$, $\omega$), the
non--potential exchange currents
${\vec j}^{\,\,a}_{5,\,\pi}(\rho\pi)$,
${\vec j}^{\,\,a}_{5,\,a_1\,\rho}(a_1)$ and the $\Delta$ excitation terms
$\vec j^{\,\,a}_{\,5,\,B}(\Delta)$ ($B=\pi$, $\rho$).
\par
The pion exchange
part of our model WAECs is similar to the one employed in
\cite{NSGK}. The representative cross sections,
presented in table I of Ref.\cite{NSGK},
are calculated using the AV18 potential \cite{AV18},
that is another high quality second generation
potential\footnote{However, its short range part is not an OBEP.}
and the $S$- and $P$-waves are taken into account in the
nucleon--nucleon final states.
\par
We also compare our results with those reported in
table I of Ref.\cite{YHH}, where the calculations were performed
(i) with the Paris potential \cite{P}; (ii)
with the currents taken in the impulse approximation; (iii)
with the $S$- and $P$-waves  taken into account in the
nucleon--nucleon final states.

\section{Numerical results} \label{sec:2}

Using the technique developed in Refs.\,\cite{JDW,ODW} one obtains
the equations for the cross sections $\sigma_{pot}(E_\nu)$ that
can be found in \cite{MRT}. The equations are the same as those of
Ref.\,\cite{NSGK}, but we treat the nucleon variables in the phase
space non-relativistically. In Ref.\,\cite{MB2}, the bounds on the
phase space are defined in the neutral channel by
\be
0\,\le\,E'_\nu\,\le\,E_\nu-\nu-2M_r+2\left[M_r(M_r\,-\,|\epsilon_B|)\right]^{\frac{1}{2}}\,,
\label{EPMAX}
\ee
\be
Max\,\left[\,-1,\,\frac{E^2_\nu+E^{\prime\,2}_\nu+4M_r(|\epsilon_B|-q_0)}{2E_\nu
E'_\nu}\,\right]\,\le\,\cos\theta\,\le\,1\,,\label{COST}
\ee
where
$M_r$ is the reduced mass of the neutron-proton system and
\mbox{$\epsilon_B$=-2.2245 MeV} is the deuteron binding energy. We
have found that it is more effective to integrate  numerically
within the bounds \be -1\,\le\,\cos\theta\,\le\,1\,,
\label{COST1} \ee \bea
0\,&\le&\,E'_\nu\,\le\,E_\nu\cos\theta\,-\,2M
\,+\,\left[\,4M^2_r
\,+\,4M_r(E_\nu \right. \nonumber \\
&&\left. \,-|\epsilon_B|)-E^2_\nu(1-\cos^2\theta) -4M_r
E_\nu\cos\theta\right]^{\frac{1}{2}}\,. \label{EPMAX1} \eea For
the charged channel, the momentum of the final lepton is
restricted by \be 0\,\le\,p_l\,\le\,p_{l,max}\,, \label{PLMAX} \ee
where $p_{l,max}$ is the solution of the equation \be
(E_\nu-p_l)^2\,+\,4M_r E(p_l)+4M_r(\Delta-E_\nu)\,=\,0\,.
\label{PLMAX} \ee Here $E(p_l)=(p^2_l+m^2_e)^\frac{1}{2}$ and \bea
\Delta &=& M_p-M_n+|\epsilon_B|\,,\,\, M_r=M_p\,,\,\, \nu\,e^-\,,
\label{NUEM}
\\
\Delta &=&
M_n-M_p+|\epsilon_B|\,,\,\,M_r=M_n\,,\,\,\bar{\nu}\,e^+\,.
\label{ANUEP} \eea  \\
We  extracted $L_{1,\,A}$ by comparing the cross section
$\sigma_{EFT}(E_\nu)$ with our cross sections
$\sigma_{pot}(E_\nu)$ using  the least square fit  and also
considering  an average value of $L_{1,\,A}$
\be
\bar{L}_{1,\,A}\,=\,\frac{\sum^N_{i=1}\,L_{1,\,A}(i)}{N}\,,
\label{LB}
\ee
where
\be
L_{1,\,A}(i)\,=\,\frac{\sigma_{pot,i}\,-\,a_i}{b_i}\,.
\label{L1AI}
\ee
We estimated the quality of the fit by the
quantity $S$ defined as
\be
S\,=\,\frac{1}{N}\,\sum^N_{i=1}\,\frac{\sigma_{EFT,i}}{\sigma_{pot,i}}\,.
\label{S}
\ee
It was found that the fit providing the average
value  (\ref{LB}) results in better agreement between
$\sigma_{EFT}(E_\nu)$ and $\sigma_{pot}(E_\nu)$ and we present the
results in table \ref{tab:1} only for this fitting procedure. We
also applied this fit to the cross sections of table I of \cite{NSGK}
(cf.\,the column NSGK).
\par
\begin{table}
\caption{ Values of the constant  $L_{1,\,A}$ obtained by the
fit to the cross sections of the reactions (\ref{NCN})-(\ref{CCA})
calculated using the NijmI, Nijm93  and OBEPQG
potentials and by the fit (NSGK) to the cross sections of table I of
Ref.\cite{NSGK}.} \label{tab:1}
\begin{tabular}{|l||c||c|c|c||c|}\hline 
reaction&  & NijmI & Nijm93 & OBEPQG & NSGK   \\\hline\hline
(\ref{NCN})&$\bar{L}_{1,\,A}$ & 4.6 & 5.2 & 4.8 &   5.4   \\
   &         S                & 1.001 & 1.001 & 1.001  &  1.000   \\\hline
(\ref{NCA})&$\bar{L}_{1,\,A}$ & 4.9 & 5.5 & 5.1 &   5.5 \\
   &         S                & 1.001 & 1.001 & 1.001  &  1.000 \\\hline
(\ref{CCN})&$\bar{L}_{1,\,A}$ & 4.1 & 5.0 & - & 6.0 \\
   &         S                & 1.001 & 1.001 & -      &  1.002 \\\hline
(\ref{CCA})&$\bar{L}_{1,\,A}$ & 4.5 & 5.4 & 6.9 &  5.6   \\
   &         S                & 1.001 & 1.000 & 0.9996 &  0.9997\\\hline
\end{tabular}
\end{table}
\par
\par
In table \ref{tab:2}, we present the scattering lengths and
the effective ranges, obtained from the NijmI, Nijm93,
OBEPQG and AV18 potentials and also the values used in the EFT
calculations \cite{MB2}. For the generation of the final state
nucleon--nucleon wave functions from the NijmI and Nijm 93
potentials, we used the program COCHASE \cite{HLS}. The program
solves the Schr\"odinger equation using the fourth--order
Runge--Kutta method. This can provide low--energy scattering
parameters  that slightly differ from those
obtained by the Nijmegen group, employing the modified Numerov
method \cite{MR}. Some refit was necessary, in order to get the
correct low--energy scattering parameters in the neutron--proton
and neutron--neutron $^{1}S_0$ states.
\par
\begin{table*}
\caption{Scattering length and effective range (in fm)
for the nucleon--nucleon system in the $^{1}S_0$ state,
corresponding to the NijmI, Nijm93 \cite{SKTS}, OBEPQG
\cite{OPT}, AV18 \cite{AV18} potentials and as used in the EFT calculations
\cite{MB2}, and their experimental values.} \label{tab:2}
\begin{center}
\begin{tabular}{|l||c|c|c|c||c||c|}\hline 
         & NijmI & Nijm93   & OBEPQG &  AV18 &  EFT  &        exp.    \\\hline\hline
$a_{np}$ & -23.72 & -23.74  & -23.74 &-23.73 &-23.7  &   -23.740$\pm$0.020$^1$  \\
$r_{np}$ &   2.65 &   2.68  &   2.73 &  2.70 &  2.70 &     2.77 $\pm$0.05$^1$ \\\hline
$a_{pp}$ &  -7.80 &  -7.79  &  -     & -7.82 & -7.82 &    -7.8063$\pm$0.0026$^2$ \\
$r_{pp}$ &   2.74 &   2.71  &  -     &  2.79 &  2.79 &     2.794$\pm$0.014$^2$  \\\hline
$a_{nn}$ & -18.16 & -18.11  & -18.10 &-18.49 &-18.5  &    -18.59$\pm$0.40$^3$   \\
$r_{nn}$ &   2.80 &   2.78  &   2.77 &  2.84 &  2.80 &      2.80$\pm$0.11$^4$   \\\hline
\end{tabular}\\
$^1$ Ref.\,\cite{CDB};\,
$^2$ Ref.\,\cite{BCSS};\,
$^3$ Ref.\,\cite{MS};\,
$^4$ Ref.\,\cite{TG}
\end{center}
\end{table*}


We shall now present the results for the reactions
(\ref{NCN})-(\ref{CCA}).
In comparing our results with \cite{MB2} we use in our calculations
their values
$G_F=1.166\times 10^{-5} GeV^{-2}$ and $g_A=-1.26$.
Instead we use the value $g_A=-1.254$, as employed
in \cite{NSGK} and \cite{YHH}, when comparing our results with
these works.
In the cross sections for the charged channel reactions (\ref{CCN})
and (\ref{CCA}) the value  $\cos \theta_C = 0.975$ is taken for
the Cabibbo angle.

\subsection{Reaction
$\nu_x\,+\,d\,\longrightarrow\,\nu'_x\,+\,n\,+\,p\,.$ }
\label{sec:21}

In table \ref{tab:3}, we present the difference in \%,
between the cross sections, obtained with the NijmI and AV18 potentials
models and the EFT cross sections, calculated with the corresponding
$\bar{L}_{1,\,A}$ from table \ref{tab:1}. Besides, we give the
differences between the cross sections,
computed with the wave functions of various potential models.\\
\begin{table}
\caption{Cross section and the differences in \% between cross sections
for the reaction (\ref{NCN}).
In the first column, $E_\nu$ [MeV] is the neutrino energy,
in the second column, $\sigma_{NijmI}$ (in $10^{-42}\times$
cm$^2$) is the cross section  calculated with the NijmI nuclear
wave functions.
Column 3 reports the differences between $\sigma_{Nijm I}$ (NijmI) and
the EFT cross section (\ref{sigB}) $\sigma_{EFT}$, calculated with the
corresponding constant $\bar{L}_{1,\,A}$ from table \ref{tab:1}. The
differences between  $\sigma_{NSGK}$ (\cite{NSGK}, table I)
and $\sigma_{EFT}$ are reported in column 4. Further,
$\Delta_{1(2)}$ is the difference  between the cross sections
$\sigma_{NijmI}$ ($\sigma_{Nijm93}$) and   $\sigma_{NSGK}$;
$\Delta_3$ is the difference between the cross sections $\sigma_{NijmI}$
and $\sigma_{YHH}$, where the cross section
$\sigma_{YHH}$ is  from (\cite{YHH}, table I).
} \label{tab:3}
\begin{tabular}{|l||c||c|c|c|c||c|c|c|}\hline
$E_\nu$ &$\sigma_{NijmI}$&NijmI&NSGK&$\Delta_1$&$\Delta_2$&$\Delta_3$
\\\hline\hline
3&0.00335&1.2&0.4&-1.1&-0.5&- \\
4&0.0306 &1.3&0.2&-0.8&-0.2&12.0 \\
5&0.0948 &1.3&0.2&-0.9&-0.2&5.0 \\
6&0.201 &1.1&0.1&-1.0&-0.3&10.2 \\
7&0.353 &1.0&0.1&-1.1&-0.4&8.1 \\
8&0.551 &1.0&0.2&-1.3&-0.5&10.1 \\
9&0.798 &1.0&0.4&-1.5&-0.7&8.9 \\
10&1.093&0.4&-0.1&-1.6&-0.8&7.6 \\
11&1.437&0.8&0.5&-1.6&-1.0&9.4 \\
12&1.831&-0.1&-0.3&-2.1&-1.2&8.5 \\
13&2.274&-0.1&0.0&-2.3&-1.4&9.9 \\
14&2.767&-0.4&0.0&-2.6&-1.7&9.5 \\
15&3.308&-0.8&-0.1&-2.9&-2.0&10.3 \\
16&3.898&-1.2&-0.3&-3.2&-2.2&9.9 \\
17&4.537&-1.6&-0.4&-3.5&-2.5&10.6 \\
18&5.223&-1.9&-0.3&-3.9&-2.9&10.3 \\
19&5.957&-2.3&-0.4&-4.2&-3.2&10.7 \\
20&6.738&-2.9&-0.6&-4.6&-3.6&10.6 \\\hline
\end{tabular}
\end{table}
Comparing the columns NijmI and NSGK of table \ref{tab:3} we can see
that the NSGK cross section is closer to the EFT cross section. This means that
the standard approach and the pionless EFT  differ, since the approximations,
made in our calculations and in EFT, coincide: the  nucleon-nucleon
final state is restricted to the $^{1}S_0$ wave and the nucleon variables
are treated non-relativistically. Besides,
the inspection of columns $\Delta_1$ and $\Delta_2$ shows that our cross
sections closely follow the NSGK cross section up to the energies when the
$P$- waves in the nucleon--nucleon final state start to contribute.
On the other hand,
as it follows from column $\Delta_3$, it is difficult to understand
the behavior of the cross section $\sigma_{YHH}$ in the whole interval
of the considered neutrino energies.

\subsection{Reaction
$\bar{\nu}_x\,+\,d\,\longrightarrow\,\bar{\nu}'_x\,+\,n\,+\,p\,.$ }
\label{sec:22}

In analogy with section \ref{sec:21}, we present in table \ref{tab:4} a
comparative analysis for the reaction (\ref{NCA}).

\begin{table}
\caption{Cross section and the differences in \% between cross sections
for the reaction (\ref{NCA}). For notations, see table \ref{tab:3},
only instead of $E_\nu$, now
$E_{\bar{\nu}}$ is the antinetrino energy in MeV. } \label{tab:4}
\begin{center}
\begin{tabular}{|l||c||c|c|c|c||c|c|c|}\hline
$E_{\bar{\nu}}$&$\sigma_{NijmI}$&NijmI&NSGK&$\Delta_1$&$\Delta_2$&$\Delta_3$\\\hline\hline
3&0.00332&0.6&0.1&-1.1&-0.5&- \\
4&0.0302 &1.0&0.2&-0.8&-0.1&9.3 \\
5&0.0928 &1.0&0.1&-0.8&-0.1&0.9 \\
6&0.196  &1.1&0.3&-0.9&-0.1&5.7 \\
7&0.342  &0.8&0.1&-1.0&-0.2&2.0 \\
8&0.531 &1.4&0.8&-1.1&-0.3&3.1 \\
9&0.765 &0.8&0.2&-1.2&-0.4&0.9 \\
10&1.043&0.6&0.2&-1.4&-0.5&-1.7 \\
11&1.364&0.1&-0.2&-1.6&-0.7&-0.7 \\
12&1.729&-0.2&-0.4&-1.7&-0.8&-2.8 \\
13&2.136&-0.3&-0.2&-1.9&-1.0&-2.1\\
14&2.585&-0.5&-0.2&-2.1&-1.2&-3.9\\
15&3.074&-0.7&-0.2&-2.4&-1.4&-4.1 \\
16&3.604&-0.9&-0.1&-2.6&-1.7&-5.6\\
17&4.173&-1.2&-0.2&-2.9&-1.9&-6.0 \\
18&4.779&-1.6&-0.3&-3.3&-2.2&-7.6 \\
19&5.422&-1.9&-0.3&-3.6&-2.5&-8.0 \\
20&6.101&-2.2&-0.2&-3.9&-2.9&-9.4 \\\hline
\end{tabular}
\end{center}
\end{table}

Clearly, our cross sections are closer to $\sigma_{EFT}$
and also to the cross section $\sigma_{NSGK}$,
 than in the neutrino-deuteron case of table \ref{tab:3}.
The behavior of the cross section
$\sigma_{YHH}$ is as little understandable as for the reaction (\ref{NCN}).
\par One can also conclude from the differences given in the columns
NijmI, NSGK, $\Delta_1$ and $\Delta_2$
of tables \ref{tab:3} and \ref{tab:4} that the cross sections for the
reactions (\ref{NCN}) and (\ref{NCA}) are described by both the potential
models and the pionless EFT with an accuracy better than 3\%.

\subsection{Reaction $\nu_e\,+\,d\,\longrightarrow\,e^-\,+\,p\,+\,p\,.$ }
\label{sec:23}

\begin{table}
\caption{Cross section and the differences in \% between cross sections
for the reaction (\ref{CCN}). For notations, see table
\ref{tab:3}. } \label{tab:5}
\begin{tabular}{|l||c||c|c|c|c||c|c|c|}\hline
$E_\nu$&$\sigma_{NijmI}$&NijmI&NSGK&$\Delta_1$&$\Delta_2$&$\Delta_3$\\\hline\hline
2&0.00338&-5.5&-0.6&-7.6&-6.7&- \\
3&0.0455 &-0.5&-0.3&-3.0&-2.0&- \\
4&0.153  & 0.5&-0.6&-1.9&-0.9&1.9 \\
5&0.340  & 1.5& 0.1&-1.6&-0.6&2.9 \\
6&0.613  & 1.9& 0.4&-1.6&-0.5&3.0 \\
7&0.978  & 1.9& 0.4&-1.6&-0.6&3.0 \\
8&1.438  & 0.0&-2.4&-1.8&-0.7&3.1 \\
9&1.997  &-0.2&-2.3&-1.9&-0.8&2.9 \\
10&2.655 & 0.1&-1.7&-2.1&-1.0&3.1 \\
11&3.415 & 3.3& 3.3&-2.4&-1.2&2.8\\
12&4.277 & 1.0& 0.3&-2.6&-1.5&2.5 \\
13&5.243 & 0.7& 0.2&-2.9&-1.8&2.4\\
14&6.311 & 0.4& 0.2&-3.2&-2.1&2.1\\
15&7.484 & 0.0& 0.2&-3.6&-2.4&1.7 \\
16&8.760 &-0.5&-0.1&-4.0&-2.8&1.4\\
17&10.14 &-0.9&-0.1&-4.4&-3.2&1.0 \\
18&11.62 &-1.3&-0.1&-4.8&-3.6&0.1 \\
19&13.21 &-1.7&-0.0&-5.3&-4.1&-0.1 \\
20&14.89 &-2.4&-0.3&-5.8&-4.5&-0.3 \\\hline
\end{tabular}
\end{table}

The comparison of the columns NijmI,  NSGK, $\Delta_1$ and $\Delta_2$
of table \ref{tab:5} shows that, disregarding the cross sections
for $E_\nu$=2 MeV, the cross sections
for the important reaction (\ref{CCN}) are  described
with an accuracy of 3.3 \%. However, while our cross sections and the cross
section \cite{NSGK} are smooth functions of the neutrino energy,
the EFT cross section exhibits abrupt changes in the region
$7\,<\,E_\nu\,<\,12$ MeV. In our opinion, the reason can be an
incorrect treatment of the Coulomb interaction between protons in the
EFT calculations. We have verified that the non-relativistic approximation for
the Fermi function, employed in \cite{MB2} is valid with a good accuracy
in the whole interval of the solar neutrino energies.
\par
Inspecting the difference of the cross sections $\Delta_3$ shows that
the cross section \cite{YHH} is of the correct size in this case.

\subsection{Reaction $\bar{\nu}_e\,+\,d\,\longrightarrow\,e^-\,+\,n\,+\,n\,.$ }
\label{sec:24}

\begin{table}
\caption{Cross section and the differences in \% between cross sections
for the reaction (\ref{CCA}). For notations, see table
\ref{tab:3}. }
\label{tab:6}
\begin{tabular}{|l||c||c|c|c|c||c|c|c|}\hline
$E_{\bar{\nu}}$&$\sigma_{NijmI}$&NijmI&NSGK&$\Delta_1$&$\Delta_2$&$\Delta_3$\\\hline\hline
5&0.0274 &-1.3&-0.9 &-2.4&-1.5&9.0 \\
6&0.116  &0.1 &-0.1 &-2.1&-1.1&8.1  \\
7&0.277  &0.2 &-0.2 &-1.8&-0.7&7.4 \\
8&0.514  &0.5 &-0.1 &-1.7&-0.6&7.1 \\
9&0.829  &0.4 &-0.2 &-1.7&-0.6&6.9  \\
10&1.224 &0.9  &0.4 &-1.7&-0.6&6.8 \\
11&1.697 &0.7  &0.2 &-1.9&-0.7&6.0\\
12&2.249 &0.6  &0.1 &-2.0&-0.8&6.1 \\
13&2.876 &0.4  &0.0 &-2.2&-1.0&5.5\\
14&3.578 &0.4  &0.2 &-2.3&-1.1&5.2\\
15&4.353 &0.0  &0.0 &-2.6&-1.3&4.9 \\
16&5.200&-0.2  &0.1 &-2.8&-1.6&4.6\\
17&6.115&-0.3  &0.2 &-3.1&-1.9&3.5 \\
18&7.097&-0.5  &0.4 &-3.4&-2.1&3.2 \\
19&8.143&-0.9  &0.2 &-3.8&-2.5&2.8 \\
20&9.251&-1.2  &0.3 &-4.1&-2.8&2.4 \\\hline
\end{tabular}
\end{table}

It follows from table \ref{tab:6} that our cross sections
for the reaction (\ref{CCA}) are in a very good agreement with
the EFT \cite{MB2} and \cite{NSGK} calculations. This confirms our
conjecture that the treatment of the Coulomb interaction between
protons \cite{MB2} in the reaction (\ref{CCN}) is not correct.
\par The calculations \cite{YHH} provide too
a small cross section. The most probable reason for this large difference
is that the Paris potential does not describe  the neutron--neutron
interaction correctly.

\section{Conclusions} \label{sec:3}

We calculated here the cross sections for the reactions of the
solar neutrinos with the deuterons, (\ref{NCN})--(\ref{CCA}),
within the standard nuclear physics approach.
We took into account the weak axial exchange currents of the OBE type,
satisfying the nuclear continuity equation (\ref{NCEt}). These currents
were constructed from the Lagrangians, possessing the chiral local
$SU(2)_L\times SU(2)_R$ symmetry, in the tree approximation.
Using the OBE potentials NijmI, Nijm93 and OBEPQG, we made
consistent calculations.
We took into account the  nucleon--nucleon interaction in the  $^{1}S_0$
final state and we treated non-relativistically the nucleon variables.
\par
Our cross sections for the reactions
(\ref{NCN}), (\ref{NCA})  and (\ref{CCA}) agree with  the EFT
cross sections \cite{MB2} and also with the cross sections
\cite{NSGK} within an accuracy better than 3 \%. The agreement for
the reaction (\ref{CCN}) is within 3.3 \%. In our opinion, the
agreement for the reaction (\ref{CCN}) can be improved by paying
more attention to the treatment of the
Coulomb interaction between the protons in the final state
in the pionless EFT calculations.

\begin{acknowledgement}
This work is supported in part by the grant GA CR 202/03/0210 and
by Ministero dell' Istruzione, dell'
Universit\`a e della Ricerca of Italy (PRIN 2003). We thank M. Rentmeester
for the correspondence.
\end{acknowledgement}

\end{document}